\documentclass[aps,pra,reprint,amsmath,amssymb,superscriptaddress,nofootinbib,longbibliography,floatfix]{revtex4-2}
\usepackage{graphicx}
\usepackage[utf8]{inputenc}
\usepackage[T1]{fontenc}
\usepackage{xcolor}
\usepackage{soul}

\usepackage{booktabs}

\usepackage{amssymb}

\usepackage{bm}
\usepackage{mathtools}

\usepackage{hyperref}
\begin{document}

	\title{Magnetocaloric performance of RE$_{5}$Pd$_2$In$_4$ (RE = Tb-Tm) compounds}

	\author{Altifani Rizky Hayyu}
	\email{altifani.hayyu@doctoral.uj.edu.pl}
	\affiliation{Jagiellonian University, Doctoral School of Exact and Natural Sciences, Faculty of Physics, Astronomy and Applied Computer Science, prof. Stanis\l{}awa \L{}ojasiewicza 11, PL-30-348 Krak\'ow, Poland}
	\affiliation{Jagiellonian University, Faculty of Physics, Astronomy and Applied Computer Science, M.~Smoluchowski Institute of Physics,
		prof. Stanis\l{}awa \L{}ojasiewicza 11, PL-30-348 Krak\'ow, Poland}	
	\author{Stanis\l{}aw Baran}
	\email{stanislaw.baran@uj.edu.pl}
	\affiliation{Jagiellonian University, Faculty of Physics, Astronomy and Applied Computer Science, M.~Smoluchowski Institute of Physics,
		prof. Stanis\l{}awa \L{}ojasiewicza 11, PL-30-348 Krak\'ow, Poland}
	\author{Andrzej Szytu\l{}a}
	\affiliation{Jagiellonian University, Faculty of Physics, Astronomy and Applied Computer Science, M.~Smoluchowski Institute of Physics,
		prof. Stanis\l{}awa \L{}ojasiewicza 11, PL-30-348 Krak\'ow, Poland}

\date{\today}
	
\begin{abstract}

Magnetocaloric performance of the RE$_{5}$Pd$_2$In$_4$ (RE = Tb-Tm) rare earth compounds has been investigated using measurements
of magnetization in the function of temperature and applied magnetic field. The maximum magnetic entropy change ($-\Delta S_{M}^{max}$) at magnetic flux density
change ($\Delta \mu_0 H$) of 0-9~T has been determined to be 3.3~J$\cdot$kg$^{-1}\cdot$K$^{-1}$ at 62~K for Tb$_{5}$Pd$_2$In$_4$,
7.0~J$\cdot$kg$^{-1}\cdot$K$^{-1}$ at 22~K for Dy$_{5}$Pd$_2$In$_4$, 12.6~J$\cdot$kg$^{-1}\cdot$K$^{-1}$ at 22~K for Ho$_{5}$Pd$_2$In$_4$, 
12.1~J$\cdot$kg$^{-1}\cdot$K$^{-1}$ at 17~K for Er$_{5}$Pd$_2$In$_4$ and 11.9~J$\cdot$kg$^{-1}\cdot$K$^{-1}$ at 9.0~K for Tm$_{5}$Pd$_2$In$_4$.
The temperature averaged entropy change (TEC) with 3~K span equals 3.2, 7.0, 12.6, 12.2 and 11.8~J$\cdot$kg$^{-1}\cdot$K$^{-1}$ for RE = Tb-Tm, respectively. The relative cooling power (RCP) and refrigerant capacity (RC) are equal to respectively
258 and 215~J$\cdot$kg$^{-1}$ in Tb$_{5}$Pd$_2$In$_4$, 498 and 325~J$\cdot$kg$^{-1}$ in Dy$_{5}$Pd$_2$In$_4$, 489 and 403~J$\cdot$kg$^{-1}$ in Ho$_{5}$Pd$_2$In$_4$,
403 and 314~J$\cdot$kg$^{-1}$ in Er$_{5}$Pd$_2$In$_4$ and 234 and 184~J$\cdot$kg$^{-1}$ in Tm$_{5}$Pd$_2$In$_4$. The magnetocaloric properties of
RE$_{5}$Pd$_2$In$_4$ are comparable to those of other known magnetocaloric materials, which show good magnetocaloric performance
at low temperatures. Among RE$_{5}$Pd$_2$In$_4$, the highest values of parameters characterizing the magnetocaloric effect are found for RE = Ho and Er.
Furthermore, for fixed RE element, the RE$_{5}$Pd$_2$In$_4$ compound displays the highest RCP and RC values when compared to those of its isostructural
RE$_{5}$T$_2$In$_4$ (T = Ni, Pt) analogues.

		\bigskip
		
		\noindent \textbf{keywords}: magnetocaloric effect, rare earth intermetallics, magnetic entropy change, relative cooling power, refrigerant capacity, temperature averaged magnetic entropy change
		
	\end{abstract}
	
	\maketitle
	
	\section{Introduction}
	\label{intro}

Magnetic materials with sizeable magnetocaloric effect (MCE) have attracted global interest due to their potential application as solid magnetic
refrigerants. The MCE is a phenomenon where certain materials undergo temperature changes in response to an externally applied
magnetic field. Among these materials, the ones containing rare earth elements with unfilled 4f subshell stand out as
prime candidates for magnetic refrigeration due to their exceptional paramagnetic susceptibility, saturation magnetization, magnetic anisotropy,
and magnetocaloric performance. The results of recent investigations can be found in both the review papers as well as in those
reporting the properties of individual compounds~\cite{D2SU00054G, 
PATINO2023414496, LI2020153810, Guo2021, BHATTACHARYYA20121239, ZHANG2009396, doi:10.1063/1.3130090, ZHANG20112602, PhysRevB.74.132405, doi:10.1063/5.0006281, Zhang_2009, DESOUZA201611, CWIK20181088, BARAN2020106837, 
LIU2022101624, MatsumotoMagnetocaloric, doi:10.1063/1.2919079, RAJIVGANDHI2018351, NIKITIN1991166, Zou_Jun-Ding_2007, WADA1999689, Burzo2010, Singh_2007, DUC2002873, samantha2006magnetocaloric_effect, C5RA06970J, MAJI2018236, 
WU2019168, Rawat_2006, R_Rawat_2001, SAHU2022103327, Zhang_2015, Mo_2015, HUO20181044, SHEN20112949, xu2012magnetocaloricHoPdAl, Sharma_2018, SHARMA2018317, SHARMA201956}.

The RE$_{5}$Pd$_2$In$_4$ (RE = Tb-Tm) intermetallics crystallize in an orthorhombic crystal structure of the Lu$_{5}$Ni$_2$In$_4$-type ($Pbam$
space group, No.~55)~\cite{Sojka200890}. RE$_{5}$Pd$_2$In$_4$, together with for example RE$_{2}$Ni$_{2}$In~\cite{Zaremba_R2Ni2In_1988,Kalychak_R2Ni2In_1990},
RE$_{5}$Ni$_{2}$In$_{4}$~\cite{zaremba1991crystal} and RE$_{11}$Ni$_{4}$In$_{9}$~\cite{PUSTOVOYCHENKO2010929} (RE = rare earth element), belongs
to a family of compounds with general composition RE$_{m+n}$T$_{2n}$X$_{m}$ (T = transition metal element), where the $m$ and $n$ refer respectively to
the numbers of the CsCl- (REIn) and AlB$_2$-related (RET$_2$) slabs the structure is composed of~\cite{sojka2008Nd11Pd4In9}. The crystal structure
of RE$_{5}$Pd$_2$In$_4$ (RE = Tb-Tm) is a special case, where within the ab-plane, the structure consists of the CsCl- and AlB$_2$-related slabs taken in the $m:n = 8:2$ ratio,
including InRE$_8$ (distorted cubes) and PdRE$_6$ (distorted trigonal prisms), forming a Z = 2, a 2:1 intergrowth variant~\cite{Zaremba2008}.
The rare earth atoms occupy three nonequivalent Wyckoff sites: the 2(a) site and two 4(g) sites with different atomic positional parameters.
The 4(g) sites are referred as 4(g)1 and 4(g)2, following the convention introduced in~\cite{Baran2021}. Along the c-axis, the layers
composed of the rare earth atoms (z = 0) are separated by layers containing the Pd and In atoms (z = $\frac{1}{2}$).

The magnetocaloric properties of RE$_{5}$Pd$_2$In$_4$ (RE = Tb-Tm) have not yet been reported, although their basic magnetic properties
are already known~\cite{Baran2021}. A ferromagnetic order sets in Tb$_{5}$Pd$_2$In$_4$ below T$_{C}=97$~K followed by appearance of
antiferromagnetic component of magnetic structure with decreasing temperature. Dy$_{5}$Pd$_2$In$_4$ orders ferromagnetically at T$_{C}=88$~K
with further transformation into ferrimagnetic structure at lower temperatures. Ho$_{5}$Pd$_2$In$_4$ is an antiferromagnet with the N\'eel
temperature T$_{N}=28.5$~K. With decreasing temperature, a second low-temperature antiferromagnetic phase develops in Ho$_{5}$Pd$_2$In$_4$ and
coexists with the high-temperature one. Er$_{5}$Pd$_2$In$_4$ is a canted antiferromagnet (T$_{N}=16.5$~K) with additional ferromagnetic components
emerging at lower temperatures. An antiferromagnetic incommensurate magnetic structure that involves two propagation vectors
($[0.073(3), 0.451(1),\frac{1}{2}]$ and $[0, 0.335(2),\frac{1}{2}]$) forms in Tm$_{5}$Pd$_2$In$_4$ below T$_{N}=4.3$~K.

Both magnetic as well as neutron diffraction measurements reveal the existence of complex magnetic structures in RE$_{5}$Pd$_2$In$_4$ (RE = Tb-Tm)
at low-temperatures accompanied by a cascade of temperature-induced magnetic transitions. Neutron diffraction data show that in the majority of the
compounds (RE = Tb-Er) the rare earth moments located at the 2(a) and 4(g)2 sites order at higher temperatures than the
moments at the 4(g)1 site. The effective magnetic moments are close to the values expected for the free RE$^{3+}$ ions, whereas the
moments in the ordered state, as derived from magnetization measurements taken at T = 2.0~K and H = 90~kOe, are significantly smaller than those
of free RE$^{3+}$ ions. The primary magnetization curves collected at 2.0~K show metamagnetic transitions, characteristic of the antiferromagnetic
component of the magnetic structure, for RE = Tb, Dy and Tm. Coercivity fields, derived from the magnetization curves at 2.0~K, decrease with
increasing number of the 4f electrons from 12.4~kOe for RE = Tb down to 0.10~kOe for RE = Tm~\cite{Baran2021}.

In this work we report for the first time the magnetocaloric performance of RE$_{5}$Pd$_2$In$_4$ (RE = Tb-Tm), which is
necessary for full characterization of magnetic properties of these compounds as well as it is important from the point of view of
possible applications in low-temperature magnetic refrigeration. The results are compared with those reported previously for the
isostructural RE$_5$Ni$_2$In$_4$ and RE$_5$Pt$_2$In$_4$ as well as for other rare earth intermetallics with good MCE performance.

\section{Materials and methods}

The samples are the same as used in our previous study~\cite{Baran2021}. They have been prepared by arc melting.
Their quality has been examined by X-ray powder diffraction at room temperature using a PANalytical X'Pert PRO diffractometer.
Detailed information on sample preparation as well as refined parameters of the crystal structure are reported in~\cite{Baran2021}.



The magnetocaloric effect in RE$_{5}$Pd$_2$In$_4$ (RE = Tb-Tm) has been investigated with the use of a Vibrating Sample Magnetometer (VSM)
option of the Physical Properties Measurement System (PPMS) by Quantum Design, equipped with a superconducting magnet running up to 9~T
(90~kOe). Before each measurement, the sample was heated to reach a paramagnetic state and then it has been demagnetized by an oscillating
magnetic field with an amplitude going to zero. Afterward, the sample was cooled down to 2.0~K in the absence of a magnetic field (zero field cooling (ZFC)
regime). Next, the final value of the magnetic field has been set. The magnetization has been measured over a temperature interval from 2.0~K
up to the one well above transition to the paramagnetic state. This procedure has been repeated for individual sample several times
for selected values of external magnetic flux density (magnetic field) ranging from 1~T (10~kOe) up to 9~T (90~kOe).

\section{Results and Discussion}

\begin{figure*}
		\centering
		\includegraphics[scale=0.25, bb=7 4 785 604]{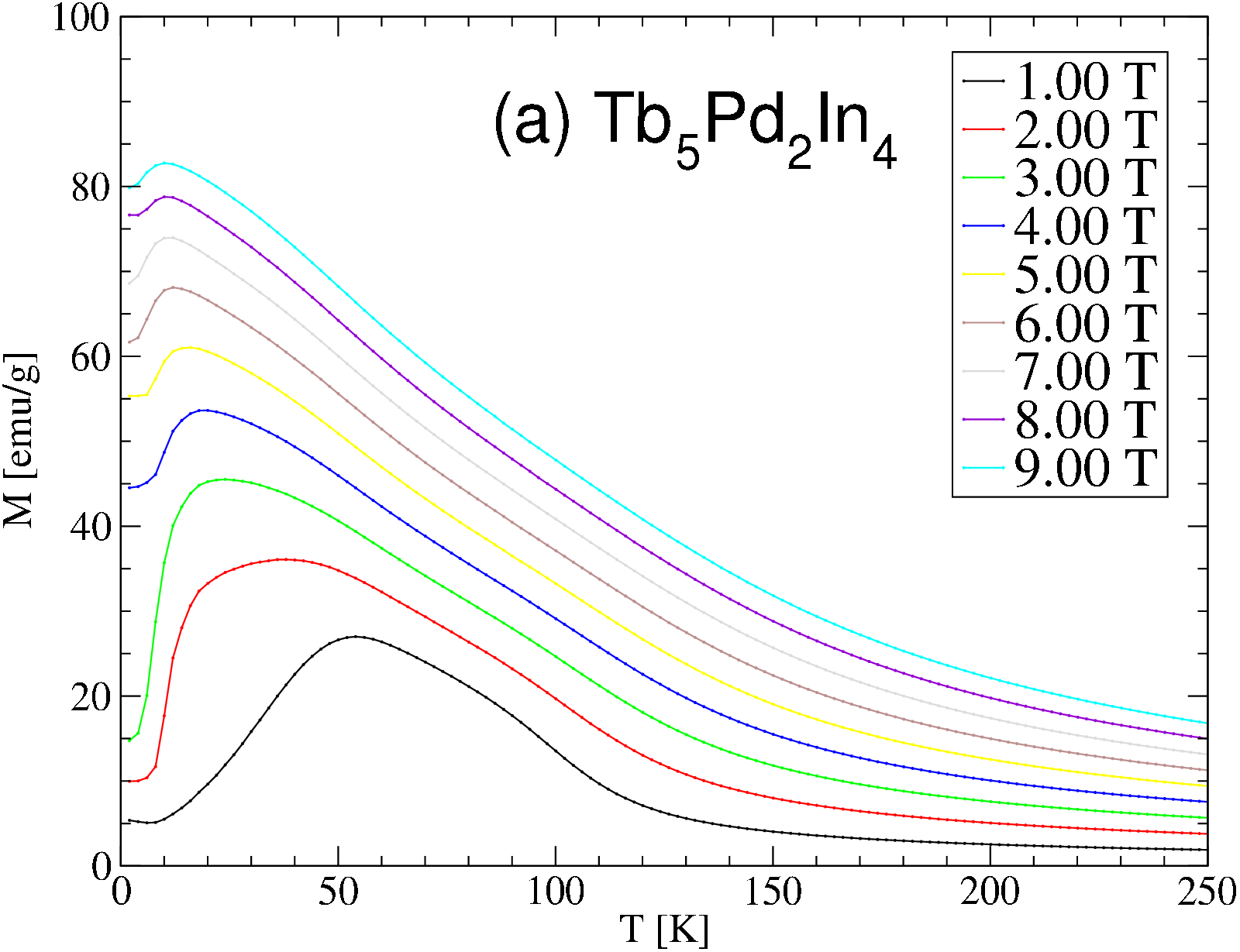}
		\includegraphics[scale=0.25, bb=7 4 785 604]{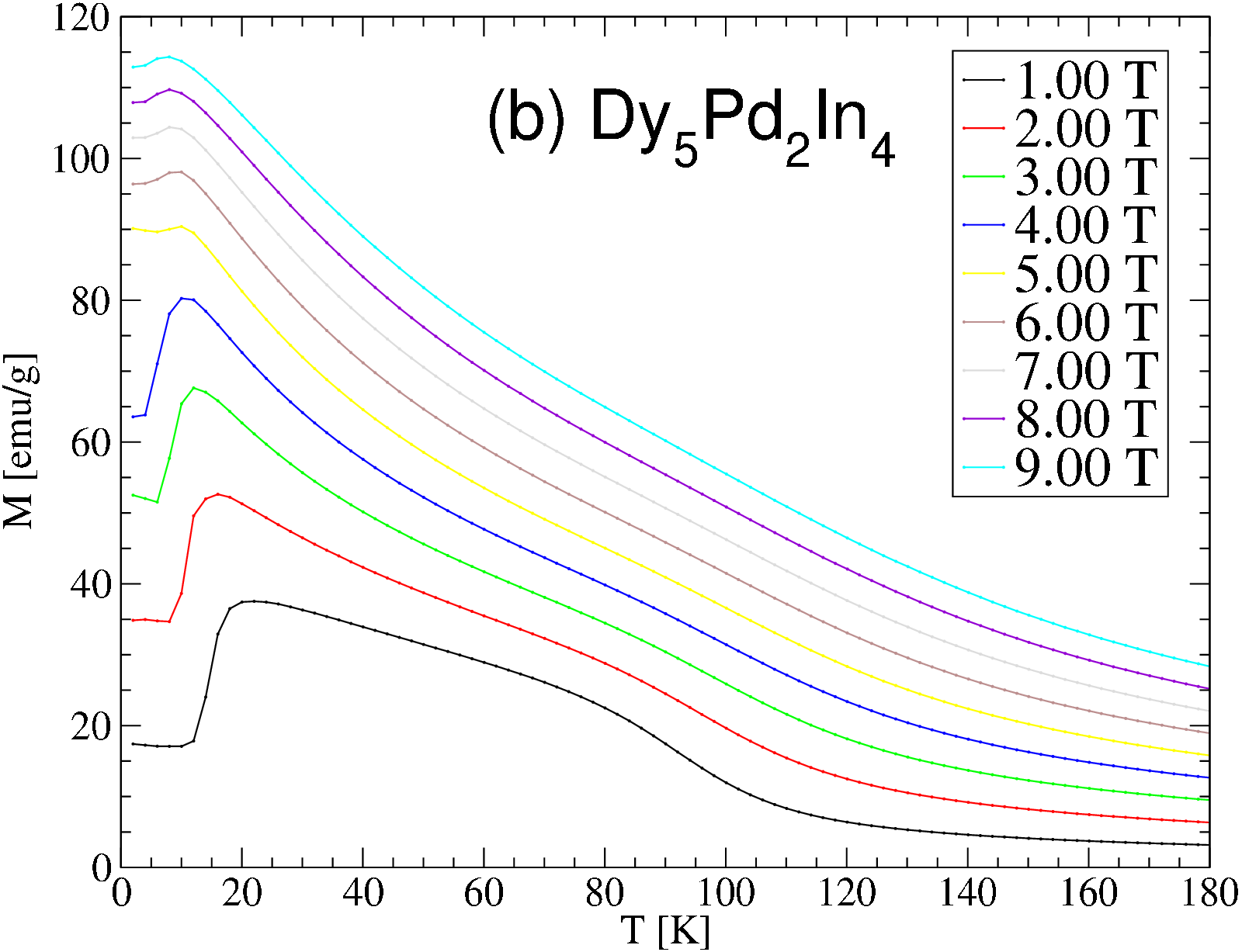}
		\includegraphics[scale=0.25, bb=7 4 785 604]{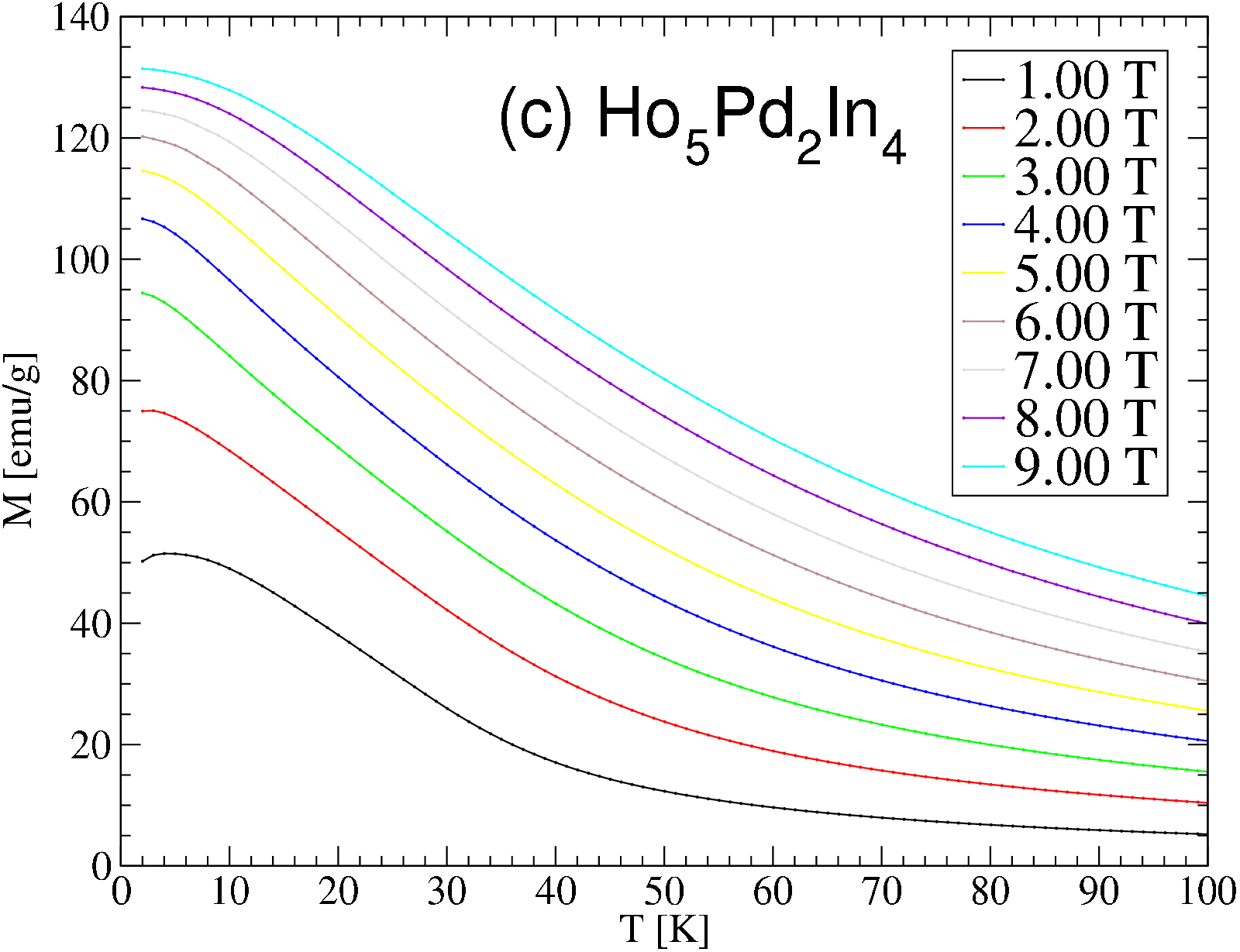}
		\includegraphics[scale=0.25, bb=7 4 785 604]{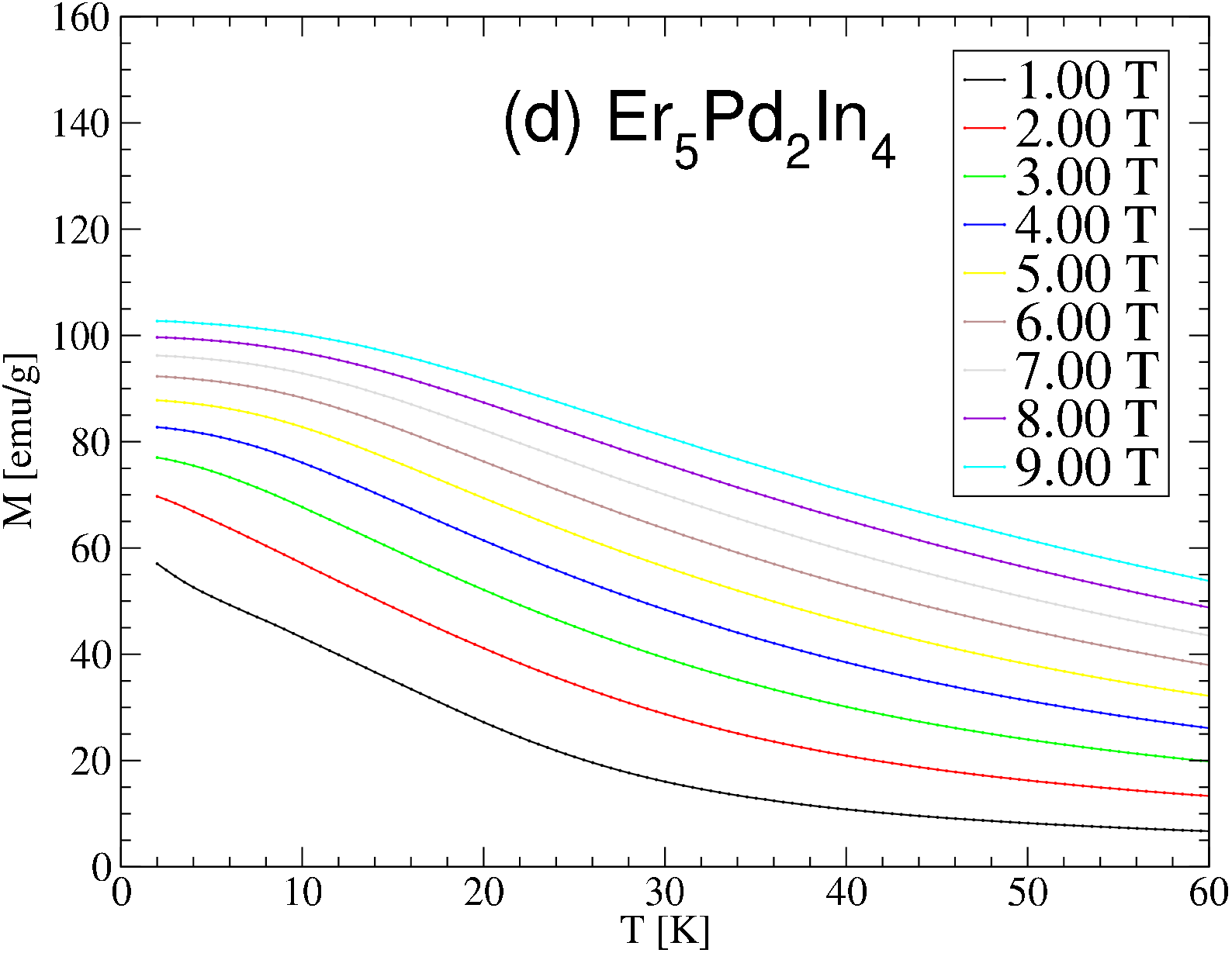}
		\includegraphics[scale=0.25, bb=7 4 785 604]{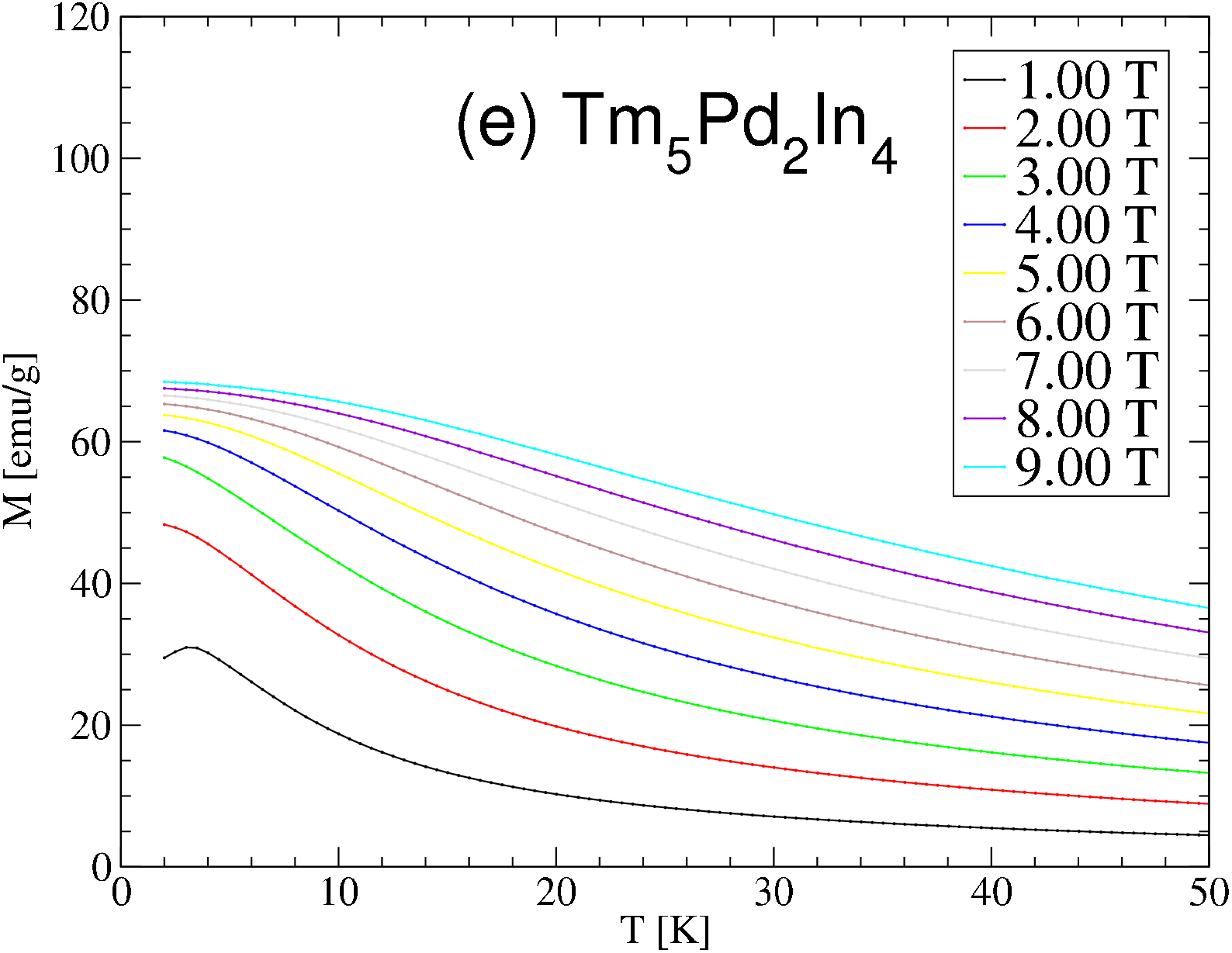}
		\caption{\label{fig:M_vs_T}ZFC magnetization vs. temperature curves under various magnetic flux density values up to 9~T for
		RE$_{5}$Pd$_2$In$_4$: (a) RE = Tb, (b) RE = Dy, (c) RE = Ho (d) RE = Er, and (e) RE = Tm.}
	\end{figure*}

Fig.~\ref{fig:M_vs_T} shows the magnetization vs. temperature M(T) curves of RE$_{5}$Pd$_2$In$_4$ (RE = Tb-Tm) collected at several fixed magnetic flux density values up to 9~T (90~kOe). All curves have an inflection point characteristic of the 
transition from ferro-/ferri- to paramagnetic state. In the case of RE = Tb and Dy, additional anomalies are visible.

To calculate the corresponding magnetic entropy changes under isothermal conditions, the following equation has been used:

\begin{equation}\Delta S_M(T,\Delta\mu_{0}H) = \int_{0}^{\mu_{0}H_{max}}  
               \left( \frac{\partial M (\mu_{0}H,T)}{\partial T} \right)_{(\mu_{0}H)}  \,\mathrm{d}\mu_{0}H
\label{eqn:dSm}
\end{equation}

\noindent where $\Delta\mu_{0}H$ represents the difference between the final ($\mu_{0}H_{f}$) and initial ($\mu_{0}H_{i}$) flux densities,
while $\left( \frac{\partial M (\mu_{0}H,T)}{\partial T} \right)_{(\mu_{0}H)}$ denotes the derivative of 
magnetization over temperature at a fixed magnetic flux density $\mu_{0}H$ ~\cite{tishin2003magnetocaloric}.

\begin{figure*}
		\centering
		\includegraphics[scale=0.25, bb=7 4 785 604]{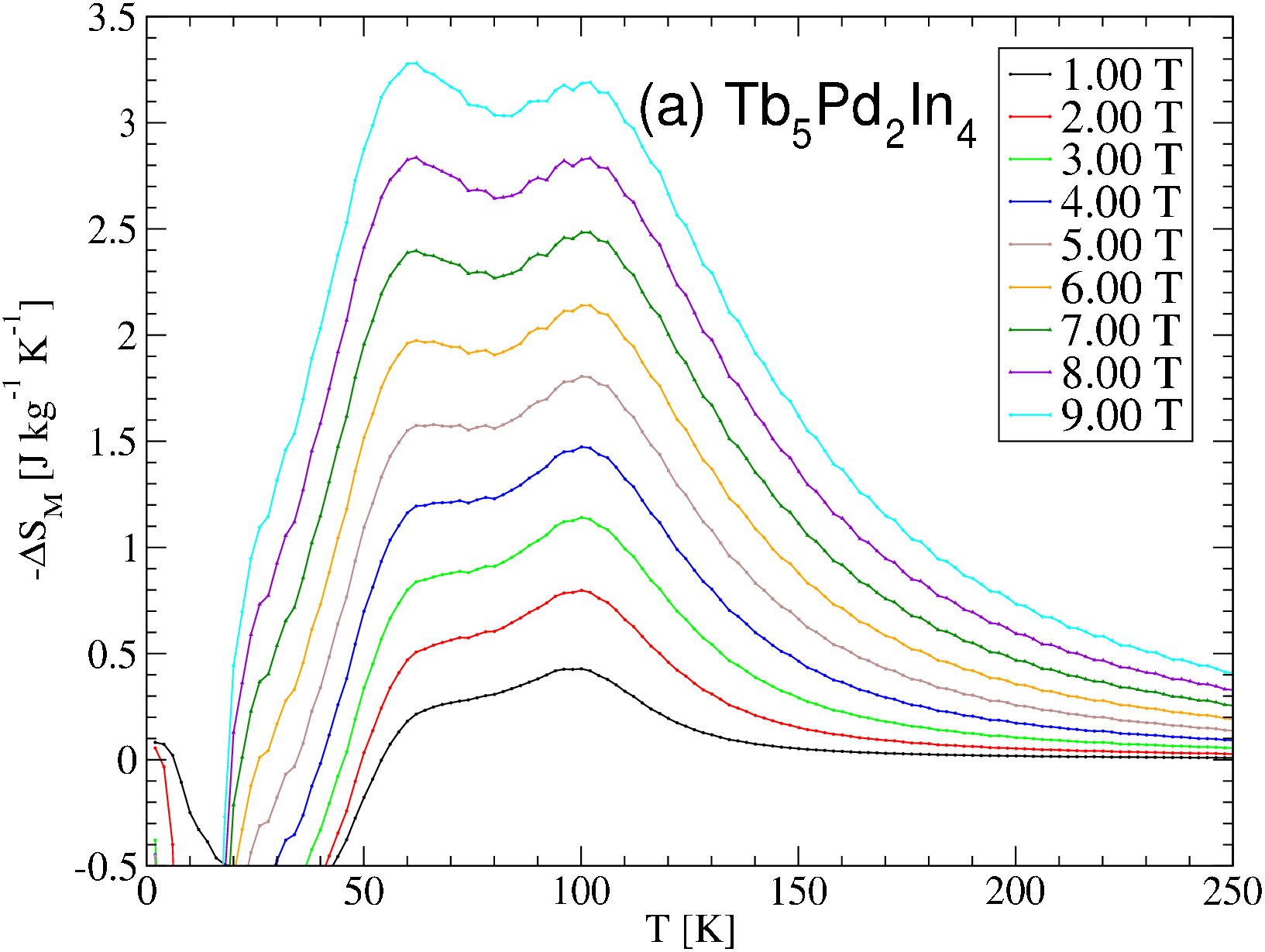} 
		\includegraphics[scale=0.25, bb=7 4 785 604]{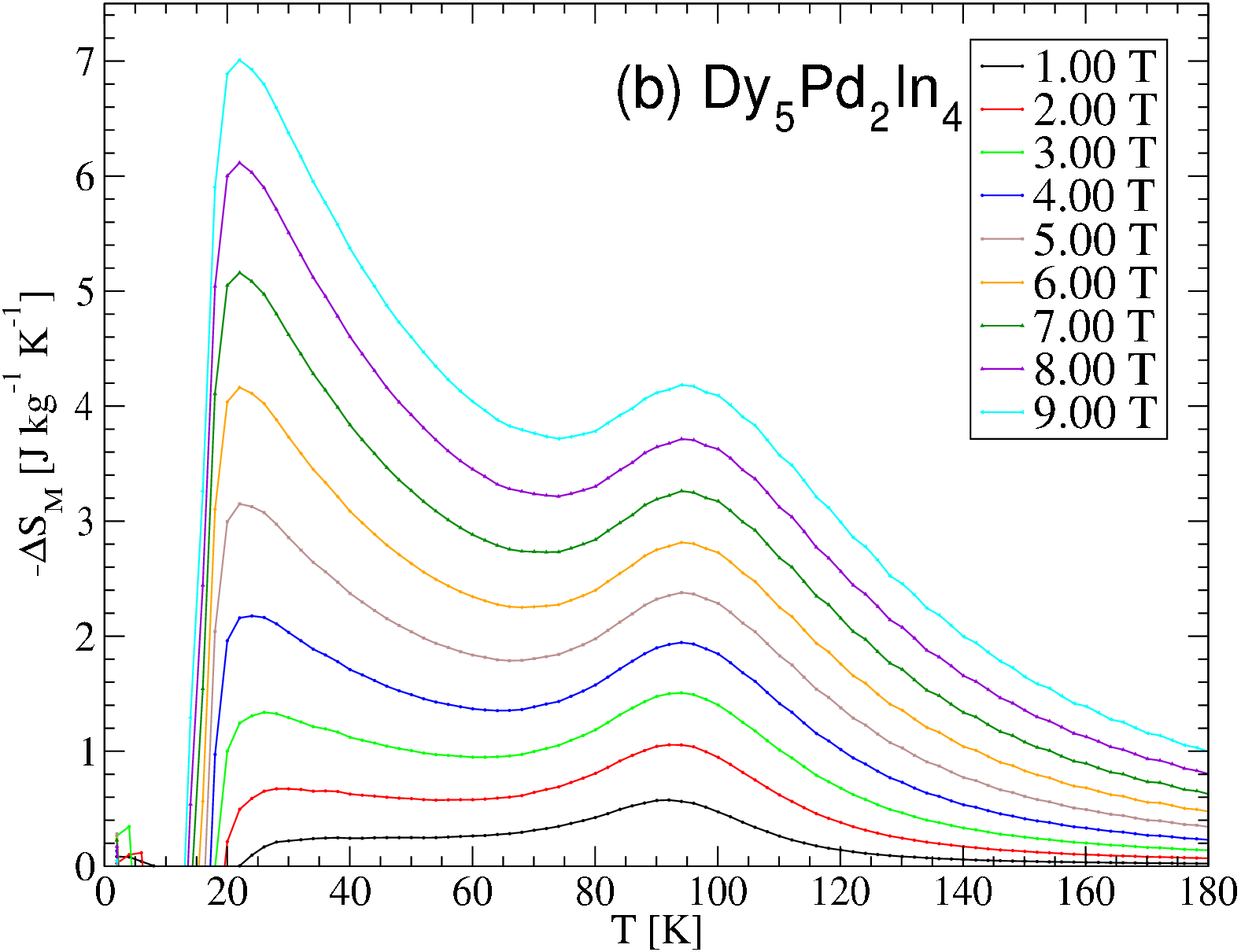}
		\includegraphics[scale=0.25, bb=7 4 785 604]{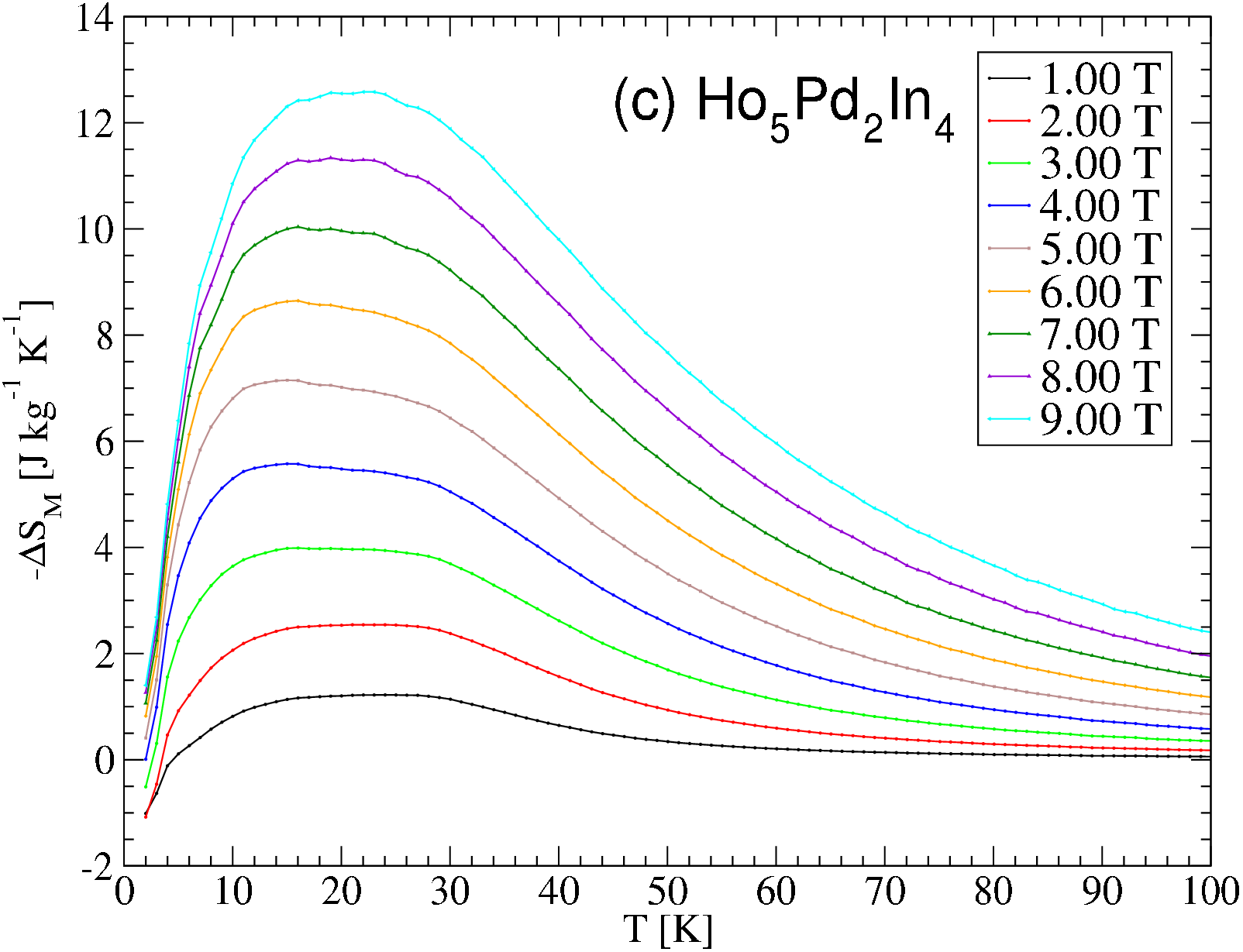} 
		\includegraphics[scale=0.25, bb=7 4 785 604]{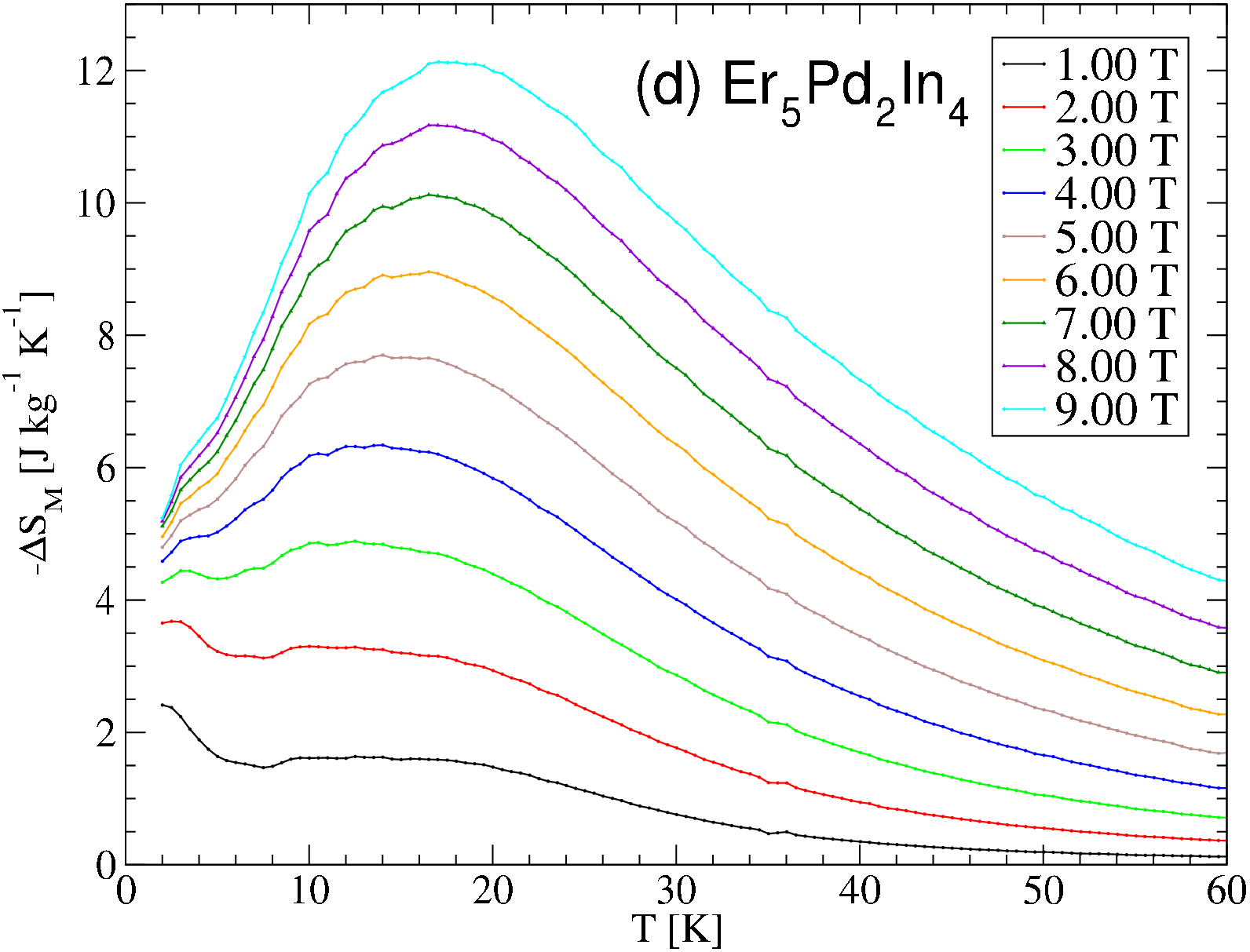}
		\includegraphics[scale=0.25, bb=7 4 785 604]{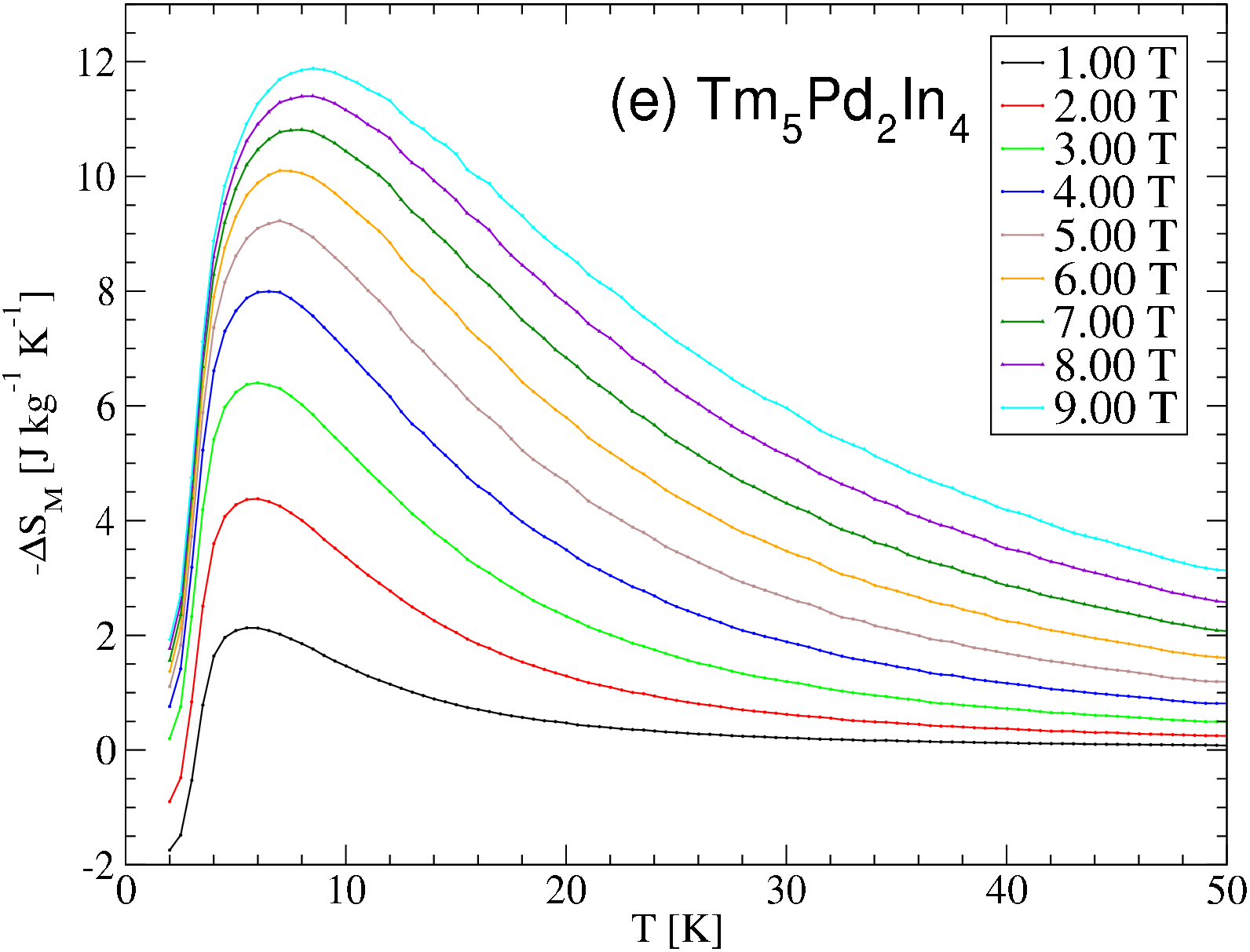} 
		\caption{\label{fig:magn_entr_vs_T}Temperature dependence of the magnetic entropy change $-\Delta S_M^{max}$, as
		derived from the $M(H,T)$ data shown in Fig.~\ref{fig:M_vs_T}, under various magnetic flux density changes $\Delta \mu_{0} H$ up to 0--9~T, for
		RE$_{5}$Pd$_2$In$_4$: (a) RE = Tb, (b) RE = Dy, (c) RE = Ho (d) RE = Er, and (e) RE = Tm.}
	\end{figure*}

Fig.~\ref{fig:magn_entr_vs_T} presents temperature dependences of the magnetic entropy change
for RE$_{5}$Pd$_2$In$_4$ (RE = Tb-Tm) as derived from the ZFC magnetization curves shown in Fig.~\ref{fig:M_vs_T}. Maximum entropy changes
$-\Delta S_M^{max}$ reach 3.3, 7.0, 12.6, 12.1 and 11.9~J$\cdot$kg$^{-1}$ under magnetic flux density change of 0-9~T for RE = Tb, Dy, Ho, Er, and Tm, respectively.
The values of $-\Delta S_M^{max}$ for other magnetic flux density changes can be found in Table~\ref{tbl:Tc_DeltaSM_RCP}. For comparison,
the table also contains $-\Delta S_M^{max}$ reported for the isostructural RE$_{5}$T$_2$In$_4$ (T = Ni, Pt) as well as for selected rare earth
intermetallics.

Two distinct maxima are found in the magnetic entropy change vs. temperature plots for Tb$_{5}$Pd$_2$In$_4$ and Dy$_{5}$Pd$_2$In$_4$
(see Figs.~\ref{fig:magn_entr_vs_T}a and ~\ref{fig:magn_entr_vs_T}b). The high-temperature maximum ($\sim$102 K for RE = Tb and $\sim$94 K for RE = Dy),
which dominates for low magnetic flux density changes, corresponds to the transition from para- to a ferromagnetic state. The low-temperature maximum
($\sim$62~K for RE = Tb and $\sim$22 K for RE = Dy), which dominates for high magnetic flux density changes, indicates an extra transformation of the
magnetic structure. The magnetic entropy changes corresponding to the low and high-temperature maxima of RE$_{5}$Pd$_2$In$_4$ (RE = Tb, Dy) are shown
separately in Figs. ~\ref{fig:magn_entr_vs_T}a and ~\ref{fig:magn_entr_vs_T}b. It is worth noting that a similar effect, involving two separate maxima in magnetic entropy change vs. temperature dependence, has also been reported for the isostructural Tb$_{5}$Pt$_2$In$_4$ and Dy$_{5}$Pt$_2$In$_4$~\cite{Hayyu2022}.

	\begin{figure*}
		\centering
		\includegraphics[scale=0.25, bb=7 4 785 604]{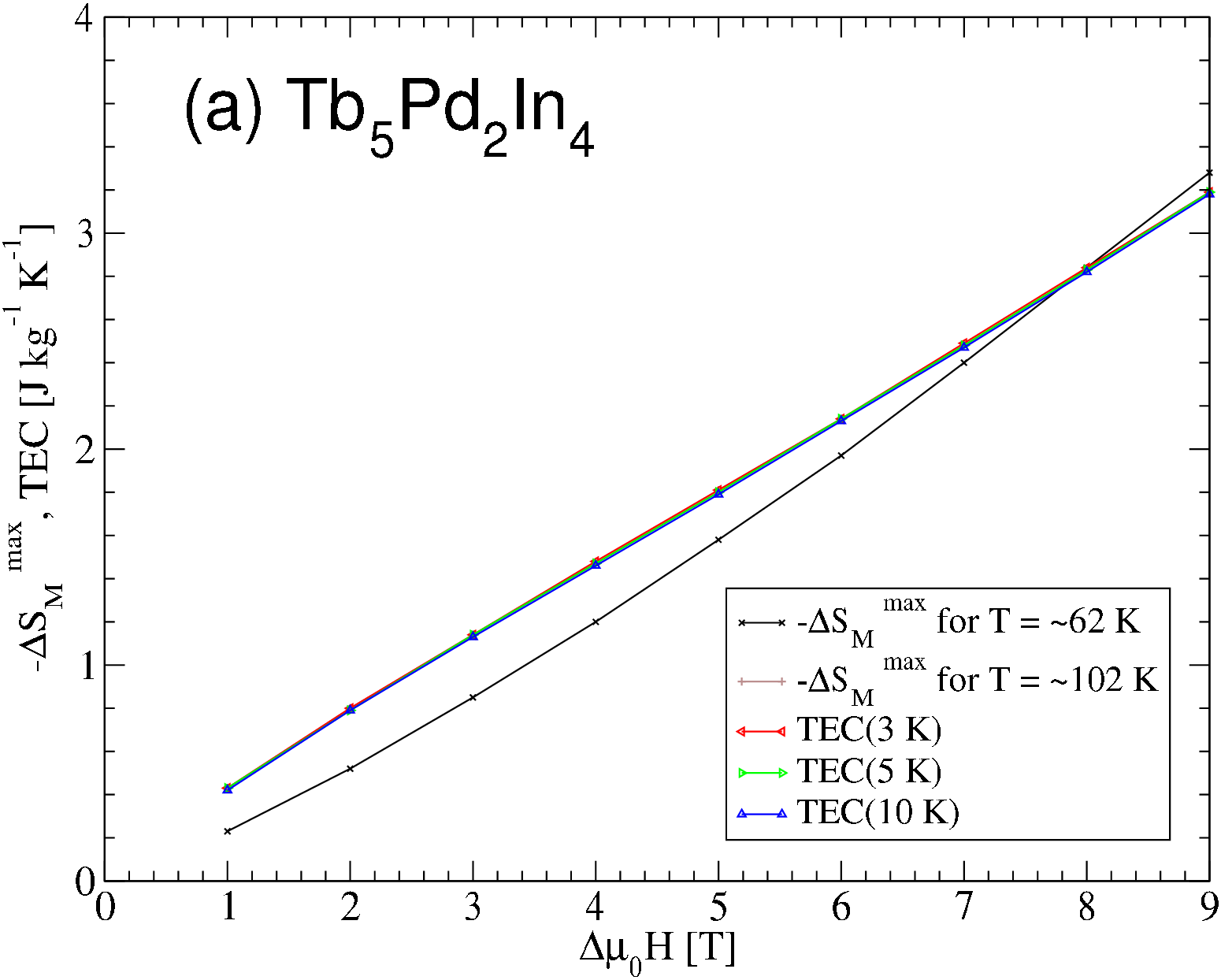}
		\includegraphics[scale=0.25, bb=7 4 785 604]{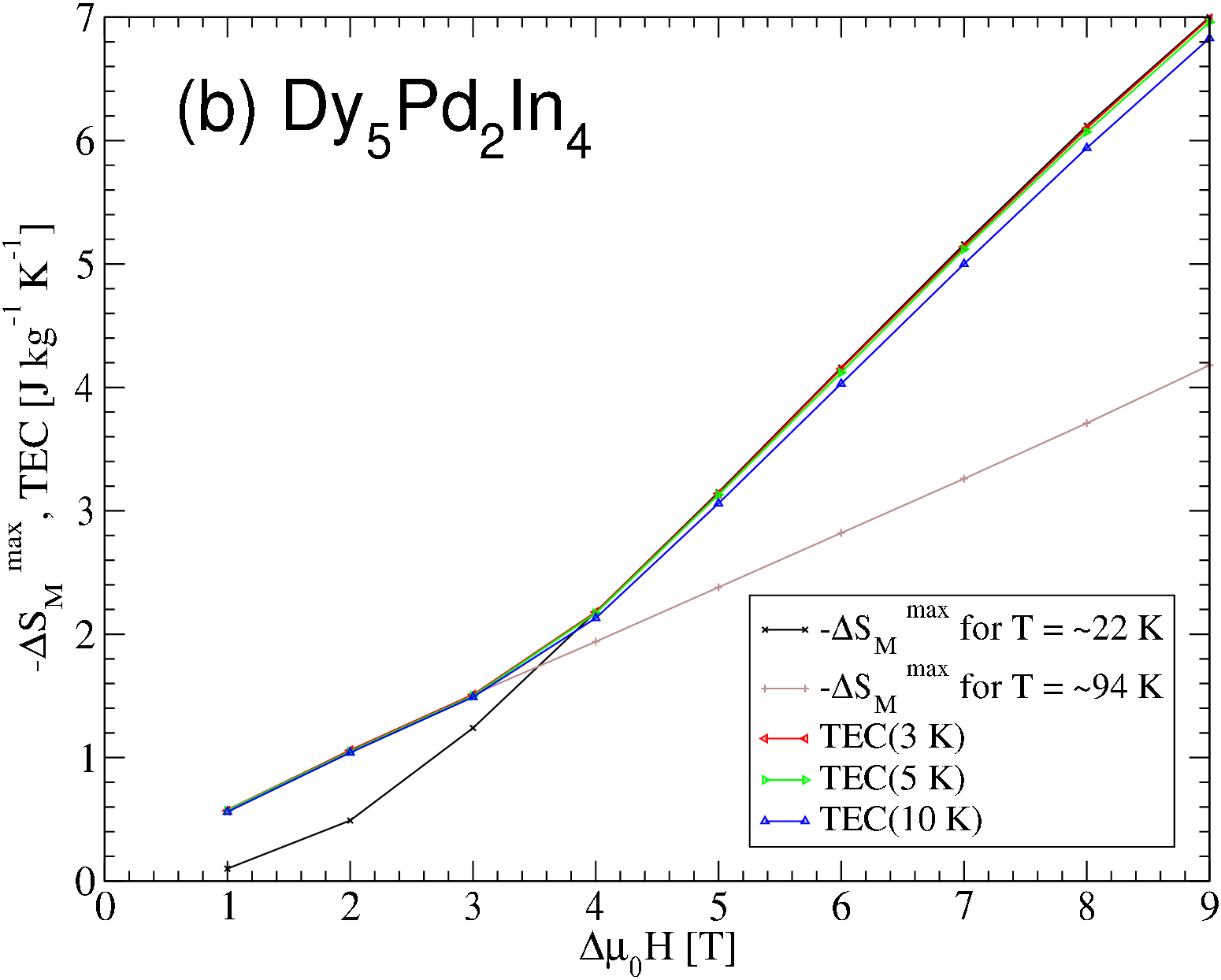}
		\includegraphics[scale=0.25, bb=7 4 785 604]{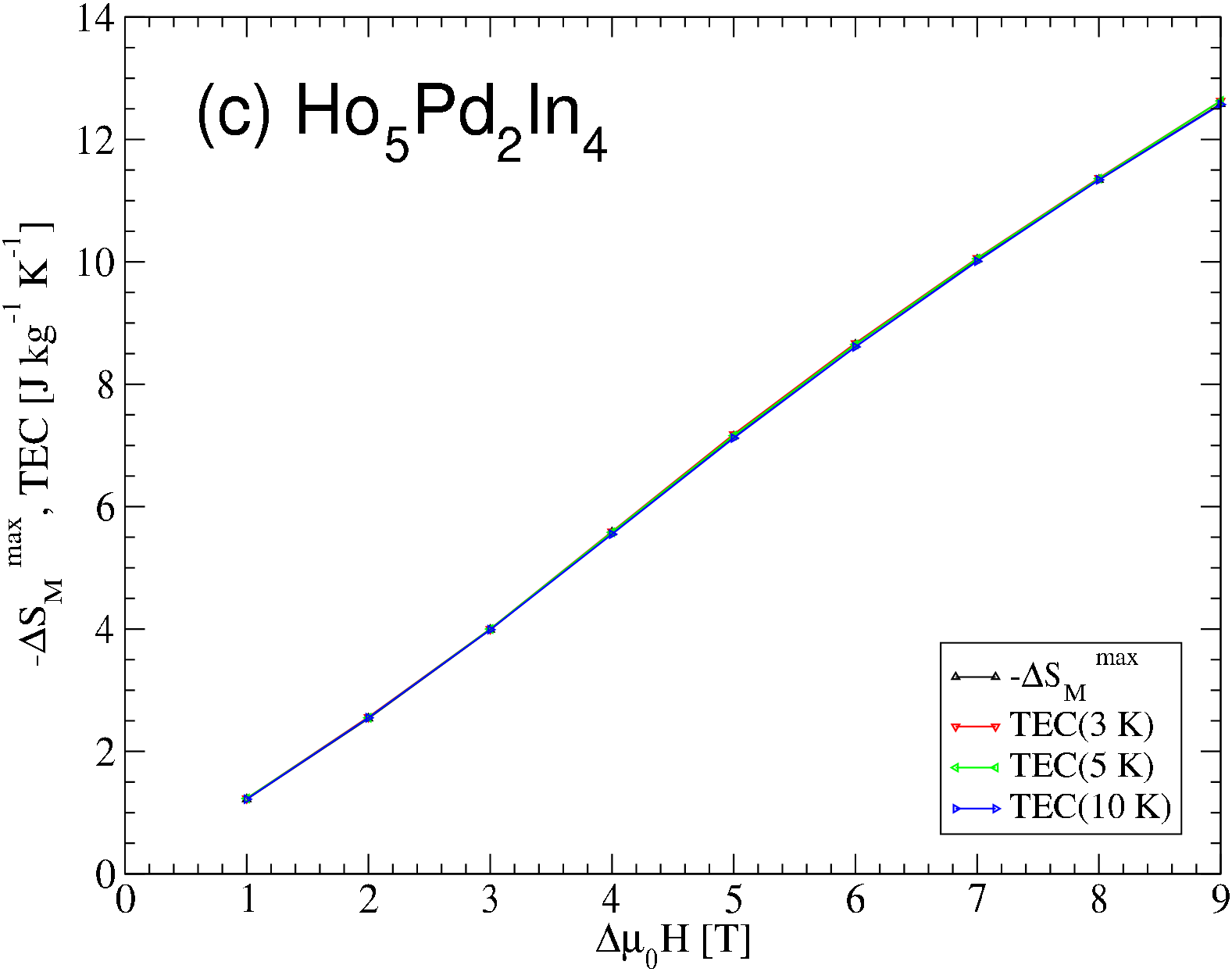}
		\includegraphics[scale=0.25, bb=7 4 785 604]{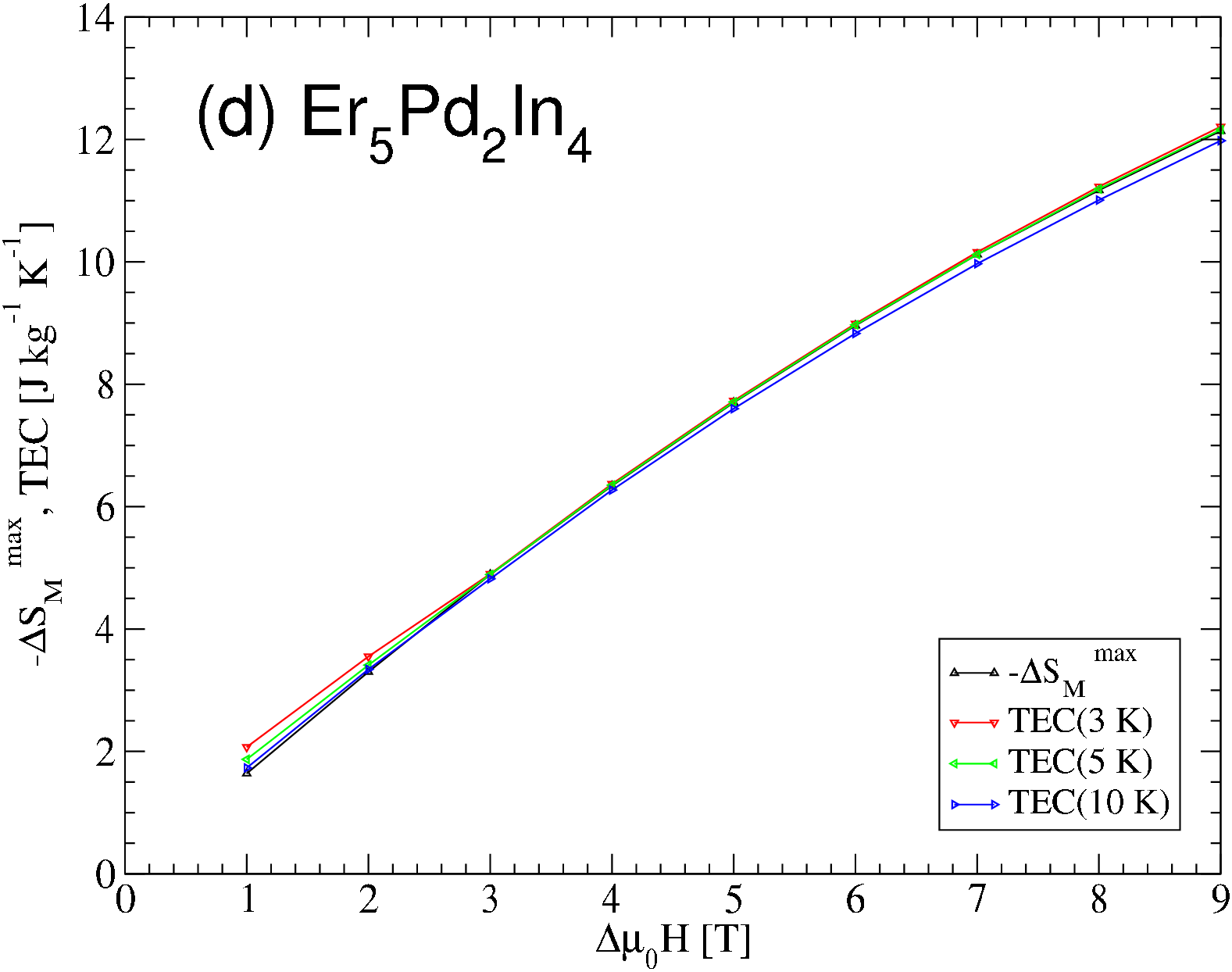}
		\includegraphics[scale=0.25, bb=7 4 785 604]{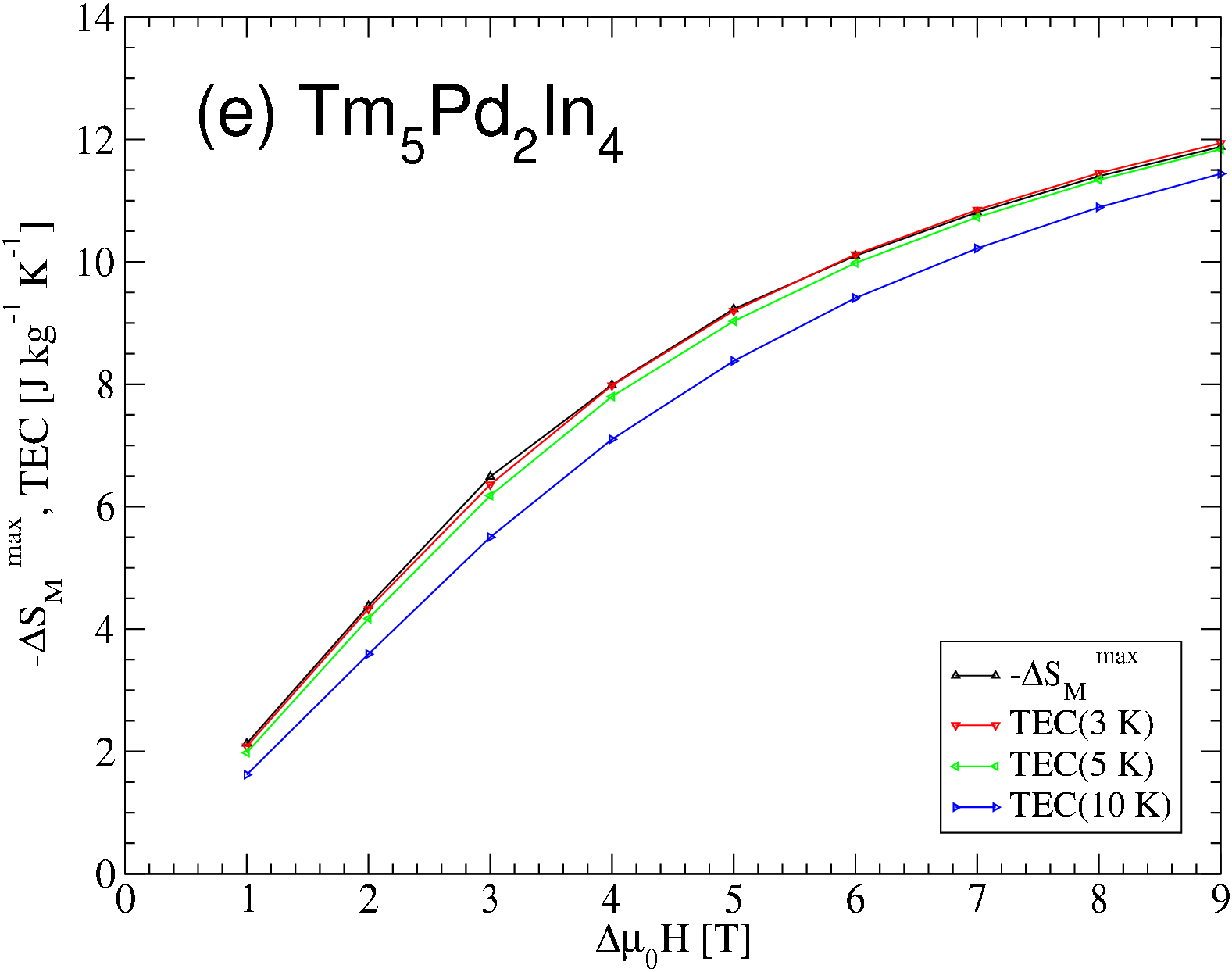}
		\caption{\label{fig:tec}The values of $-\Delta S_M^{max}$, TEC(3 K), TEC(5 K), and TEC(10 K) plotted against
		magnetic flux density change $\Delta \mu_{0}H$ for RE$_{5}$Pd$_2$In$_4$: (a) RE = Tb, (b) RE = Dy,
		(c) RE = Ho, (d) RE = Er, and (e) RE = Tm, respectively.}
	\end{figure*}
	

The temperature-averaged magnetic entropy change (TEC) is another parameter used for characterizing a magnetocaloric material. TEC is defined by the following formula ~\cite{griffith2018material-based}:

\begin{align} 
\MoveEqLeft TEC(\Delta T_\mathrm{lift},\Delta\mu_{0}H) =
\nonumber \\  &\frac{1}{\Delta T_\mathrm{lift}} \max_{T_\mathrm{mid}}
  \Bigg\{\int_{T_\mathrm{mid}-\frac{\Delta T_\mathrm{lift}}{2}}^{T_\mathrm{mid}+\frac{\Delta T_\mathrm{lift}}{2}} \Delta S_{\mathrm{M}}(T,\Delta\mu_{0}H)\,\mathrm{d}T\Bigg\}
\label{eqn:tec}
\end{align}

\noindent where $T_\mathrm{mid}$ is the center temperature of the temperature span $\Delta T_\mathrm{lift}$. The value of $T_\mathrm{mid}$ is determined by finding the one that maximizes the integral appearing in Eq.~\ref{eqn:tec}. 



The TEC values of RE$_{5}$Pd$_2$In$_4$ (RE = Tb–Tm), calculated for the temperature spans of 3, 5 and 10 K, are presented in Fig.~\ref{fig:tec}.
Within the 3~K span and magnetic flux density change of 0-9~T, TEC reaches 3.2~J$\cdot$kg$^{-1}\cdot$K$^{-1}$ (RE = Tb),
7.0~J$\cdot$kg$^{-1}\cdot$K$^{-1}$ (RE = Dy), 12.6~J$\cdot$kg$^{-1}\cdot$K$^{-1}$ (RE = Ho), 12.2~J$\cdot$kg$^{-1}\cdot$K$^{-1}$ (RE = Er) and 11.8~J$\cdot$kg$^{-1}\cdot$K$^{-1}$ (RE = Tm), respectively.
TEC is slightly decreased with increasing temperature span T$_{lift}$.
The values of TEC are close to the ones of corresponding $-\Delta S_M^{max}$,
indicating that magnetocaloric performance is maintained over temperature integral of at least 10~K. It is noticeable that for RE = Tb and Dy
the TEC values are always related to the high-temperature maximum for the low magnetic flux density changes and the low-temperature
one for the high magnetic flux density changes (see Figs.~\ref{fig:tec}a and ~\ref{fig:tec}b).


Besides $\Delta S_M^{max}$ and TEC, the relative cooling power (RCP)~\cite{gschneidner_pecharsky2000magnetocaloric_materials} and
refrigerant capacity (RC)~\cite{wood_potter1985general_analysis} are also used to characterize magnetocaloric
performance. The latter parameters correspond to the amount of heat transfer between the cold and hot reservoirs in an 
ideal refrigeration cycle~\cite{Li_2016}. RCP is defined as the product of the maximum magnetic entropy change $-\Delta S_M^{max}$ and the full width 
at half maximum $\delta T_{FWHM}$ in the -$\Delta S_M(T)$ curve by the equation:

\begin{equation}
 RCP = -\Delta S_M^{max} \times \delta T_{FWHM}
\label{eqn:rcp}
\end{equation}

RC is calculated by numerically integrating the area under the {-$\Delta S_M(T)$} curve, while taking the half maximum of the peak as the integration limits:

\begin{equation}
 RC = \int_{T_{1}}^{T_{2}} |\Delta S_M| \,\mathrm{d}T
\label{eqn:rc}
\end{equation} 

\noindent where $T_1$ and $T_2$ are the lower and upper limits of the FWHM temperature range, respectively.

\begin{table*}
\begin{footnotesize}
  \begin{flushleft}
  \normalsize
\caption{\label{tbl:Tc_DeltaSM_RCP}\ The temperatures of maximum entropy change, maximum entropy changes $-\Delta S_M^{max}$
together with RCP and RC values under magnetic flux density changes $\Delta \mu_0 H$ = 0--2 T, 0--5 T, 0--7 T and 0--9 T for the RE$_{5}$Pd$_2$In$_4$ (RE 
= Tb-Tm), isostructural RE$_{5}$T$_2$In$_4$ (T = Ni, Pt) and selected rare earth intermetallics.\\}
  \end{flushleft}
  \vspace{-0.2 cm}
  \footnotesize
  \begin{tabular*}{0.99\textwidth}{@{\extracolsep{\fill}}lllllllllllllll}
    \hline
    Materials & Temp. for $-\Delta S_{\mathrm{M}}^{\mathrm{max}}$ [K] & \multicolumn{4}{c}{$-\Delta S_{\mathrm{M}}^{\mathrm{max}}$[J$\cdot$kg$^{-1}\cdot$K$^{-1}$]} & \multicolumn{4}{c}{RCP [J$\cdot$kg$^{-1}$]} & \multicolumn{4}{c}{RC [J$\cdot$kg$^{-1}$]} & Ref.\\
\cline{3-6} \cline{7-10} \cline{11-14}  
\multicolumn{1}{l}{} & \multicolumn{1}{l}{} & \multicolumn{1}{l}{0--2 T} & \multicolumn{1}{l}{0--5 T} & \multicolumn{1}{l}{0--7 T} & \multicolumn{1}{l}{0--9 T} &
	\multicolumn{1}{l}{0--2 T} & \multicolumn{1}{l}{0--5 T} & \multicolumn{1}{l}{0--7 T} & \multicolumn{1}{l}{0--9 T} & \multicolumn{1}{l}{0--2 T} & \multicolumn{1}{l}{0--5 T} & \multicolumn{1}{l}{0--7 T} & \multicolumn{1}{l}{0--9 T} & \\
    \hline
 Dy$_{5}$Ni$_2$In$_4$ & 103 & 1.8 & 3.6 & 4.7 & - & 49 & 178 & 286 & - & 37 & 130 & 209 & - & \citep{zhang2018investigation}\\
 Ho$_{5}$Ni$_2$In$_4$ & 19 & 2.6 & 7.1 & 10.1 & - & 84 & 298 & 458 & - & 66 & 234 & 352 & - & \citep{zhang2018investigation}\\
 Er$_{5}$Ni$_2$In$_4$ & 20 & 3.3 & 7.7 & 10.2 & - & 71 & 248 & 377 & - & 52 & 180 & 273 & - & \citep{zhang2018investigation}\\
 Gd$_{5}$Pt$_2$In$_4$ & 78 & 1.0 & 2.2 & 3.0 & 3.7 & 58 & 172 & 261 & 359 & 48 & 139 & 209 & 290 & \cite{Hayyu2022}\\
 Tb$_{5}$Pt$_2$In$_4$ & 45, 110 & 0.8 & 1.7 & 2.5 & 3.4 & 82 & 198 & 303 & 428 & 57 & 165 & 248 & 340 & \cite{Hayyu2022}\\
 Dy$_{5}$Pt$_2$In$_4$ & 25, 95 & 1.1 & 2.9 & 4.7 & 6.3 & 45 & 290 & 250 & 363 & 33 & 201 & 180 & 263 & \cite{Hayyu2022}\\
 Ho$_{5}$Pt$_2$In$_4$ & 23 & 2.5 & 6.9 & 9.5 & 11.8 & 94 & 302 & 451 & 607 & 82 & 254 & 373 & 495 & \cite{Hayyu2022}\\
 Er$_{5}$Pt$_2$In$_4$ & 14 & 3.6 & 7.5 & 9.6 & 11.4 & 79 & 218 & 328 & 434 & 63 & 175 & 256 & 341 & \cite{Hayyu2022}\\
 Tm$_{5}$Pt$_2$In$_4$ & 8 & 3.2 & 7.7 & 9.2 & 10.2 & 34 & 125 & 189 & 260 & 27 & 98 & 150 & 205 & \cite{Hayyu2022}\\
 Tb$_{5}$Pd$_2$In$_4$ & 62, 102 & 0.8 & 1.8 & 2.5 & 3.3 & 53 & 161 & 258 & 377 & 44 & 133 & 215 & 312 & this work\\
 Dy$_{5}$Pd$_2$In$_4$ & 22, 94 & 1.1 & 3.2 & 5.2 & 7.0 & 100 & 314 & 498 & 672 & 65 & 216 & 325 & 406 & this work\\
 Ho$_{5}$Pd$_2$In$_4$ & 22 & 2.5 & 7.2 & 10.0 & 12.6 & 94 & 326 & 489 & 661 & 81 & 269 & 403 & 541 & this work\\
 Er$_{5}$Pd$_2$In$_4$ & 17 & 3.3 & 7.7 & 10.1 & 12.1 & 103 & 287 & 403 & 528 & 91 & 227 & 314 & 409 & this work\\
 Tm$_{5}$Pd$_2$In$_4$ & 9 & 4.4 & 9.2 & 10.8 & 11.9 & 85 & 156 & 234 & 320 & 50 & 123 & 184 & 249 & this work\\
 Gd$_{11}$Co$_4$In$_9$ & 85 & 5.0 & 9.0 & 11.0 & - & 106 & 358 & - & - & 82 & 270 & - & - & \cite{zhang2020structural}\\
 Tb$_{11}$Co$_4$In$_9$ & 42 & 1.2 & 3.3 & 4.4 & 5.5 & 45 & 180 & - & 522 & 35 & 140 & - & 391 & \cite{Baran_R11Co4In9_Phase_Trans}\\
 Dy$_{11}$Co$_4$In$_9$ & 35 & 1.1 & 3.5 & 4.7 & - & 27 & 128 & - & - & 20 & 98 & - & - & \cite{zhang2020structural}\\
 Ho$_{11}$Co$_4$In$_9$ & 20 & 3.0 & 9.2 & 12.3 & - & 87 & 307 & - & - & 66 & 229 & - & - & \cite{zhang2020structural}\\
 Er$_{11}$Co$_4$In$_9$ & 12 & 6.0 & 10.9 & 12.8 & 14.3 & 85 & 265 & - & 605 & 66 & 205 & - & 463 & \cite{Baran_R11Co4In9_Phase_Trans}\\
 Gd$_{11}$Ni$_4$In$_9$ & 98 & 1.4 & 2.9 & 3.6 & - & 54 & 171 & 269 & - & 38 & 131 & 206 & - & \cite{ZHANG2021155863}\\
 Dy$_{11}$Ni$_4$In$_9$ & 19, 92 & 0.8 & 3.8 & 6.0 & - & 21 & 109 & 195 & - & 16 & 81 & 145 & - & \cite{ZHANG2021155863}\\
 Ho$_{11}$Ni$_4$In$_9$ & 16 & 1.8 & 8.8 & 12.4 & - & 32 & 198 & 353 & - & 24 & 152 & 269 & - & \cite{ZHANG2021155863}\\
 Gd$_{6}$Co$_{2.2}$In$_{0.8}$ & 42, 78 & 2.6 & 8.4 & 11.8 & - & 148 & 532 & 814 & - & 97 & 381 & 634 & - & \cite{ZHANG2021107254}\\
 Tb$_{6}$Co$_{2.2}$In$_{0.8}$ & 38 & 1.4 & 5.9 & 9.0 & - & 39 & 214 & 394 & - & 28 & 157 & 284 & - & \cite{ZHANG2021107254}\\
 Dy$_{6}$Co$_{2.2}$In$_{0.8}$ & 53 & 1.5 & 6.1 & 9.6 & - & 36 & 288 & 517 & - & 25 & 211 & 390 & - & \cite{ZHANG2021107254}\\
 Ho$_{6}$Co$_{2.2}$In$_{0.8}$ & 19 & 3.8 & 15.4 & 20.8 & - & 74 & 370 & 626 & - & 57 & 279 & 466 & - & \cite{ZHANG2021107254}\\
 Gd$_{2}$In & 194 & 2.8 & - & - & - & - & - & - & - & - & - & - & - & \citep{BHATTACHARYYA20121239}\\ 
 Tb$_{2}$In & 165 & 3.5 & 6.6 & - & - & 200 & 660 & - & - & - & - & - & - &\citep{ZHANG2009396}\\
 Dy$_{2}$In & 130 & 4.6 & 9.2 & - & - & 230 & 736 & - & - & - & 545 & - & - &\citep{Zhang_2009}\\
 Dy$_{2}$In & 125 & - & - & 8.8 & - & - & - & - & - & - & - & - & - & \citep{YAO201937}\\
 Ho$_{2}$In & 85 & 5.0 & 11.2 & - & - & 125 & 560 & - & - & - & 360 & - & - & \citep{doi:10.1063/1.3130090, ZHANG2009396}\\
 Er$_{2}$In & 46 & 7.9 & 16.0 & - & - & - & - & - & - & - & 490 & - & - & \citep{ZHANG20112602}\\ 
 GdPt$_{2}$ & 28 & 3.2$^{a}$ & - & 11.3 & - & - & - & - & - & - & - & - & - & \citep{PhysRevB.74.132405}\\  
 DyPt$_{2}$ & 10 & 5.0 & 9.4 & 21.2$^{b}$ & - & - & - & - & - & 50 & 150 & - & - &  \citep{doi:10.1063/1.3253729}\\ 
    \hline
  \end{tabular*}
\end{footnotesize}
  \\
$^a$ $\Delta \mu_0 H$ = 0--1 T | $^b$ $\Delta \mu_0 H$ = 0--8 T\\
\end{table*}


Table~\ref{tbl:Tc_DeltaSM_RCP} shows the magnetocaloric performance of RE$_{5}$Pd$_2$In$_4$ (RE = Tb-Tm) in comparison
with isostructural RE$_{5}$T$_2$In$_4$ (T = Ni, Pt) and other selected rare-earth-based intermetallics in terms of temperature at which
$-\Delta S_{\mathrm{M}}^{\mathrm{max}}$ is reached as well as the values of $-\Delta S_{\mathrm{M}}^{\mathrm{max}}$, RCP and RC.



For a selected rare earth element, the MCE performance of a member of the RE$_{5}$Pd$_2$In$_4$ (RE = Tb-Tm) family of compounds shows
similar performance to that of the isostructural RE$_{5}$T$_2$In$_4$ (T = Ni, Pt) as well as to that of its
RE$_{11}$T$_4$In$_9$ (T = Co, Ni) analogues. Much more effected is the temperature at which $-\Delta S_{\mathrm{M}}^{\mathrm{max}}$
occurs. For example, for Er$_{5}$T$_2$In$_4$ such temperature varies from 14~K for T~=~Pt up to 20~K for T~=~Ni. Therefore, the selection of transition metal (T) and stoichiometry enables us to get a compound that shows maximum magnetocaloric performance
at a certain temperature. Moreover, combining different compounds allows to obtain hybrid materials that show good
magnetocaloric performance over desired temperature interval. This feature is important from the point of view of
application in low-temperature refrigeration.

It is worth noting that RE$_{5}$Pd$_2$In$_4$ (RE = Tb-Tm) have higher values of RCP and RC when compared to their RE$_{5}$T$_2$In$_4$
analogues with T = Ni or Pt. Among RE$_{5}$Pd$_2$In$_4$, the best magnetocaloric performance is found for RE = Ho and Er, making
these two compounds competitive to known rare-earth-based magnetocaloric materials, which show good magnetocaloric performance
at low temperatures.

\section{Conclusions}

The magnetocaloric effect in RE$_{5}$Pd$_2$In$_4$ (RE = Tb-Tm) has been investigated using magnetometric measurements
in the function of temperature and applied magnetic field. Based on these data, $-\Delta S_{M}^{max}$, TEC, RC and RCP values have been
determined and compared to those of selected ternary rare-earth-based intermetallics. The maximum of $-\Delta S_{M}^{max}$ is reached at
62~K, 22~K, 22~K, 17~K, and 9~K for RE = Tb, Dy, Ho, Er, and Tm, respectively. For selected rare earth elements, the RE$_{5}$Pd$_2$In$_4$
compound shows the highest RCP and RC values when compared to those of its RE$_{5}$T$_2$In$_4$ (T = Ni, Pt) isostructural
analogues~\cite{zhang2018investigation,Hayyu2022}.

The magnetocaloric performance of RE$_{5}$Pd$_2$In$_4$ (RE = Tb-Tm) is comparable to that found for other magnetocaloric materials
showing remarkable magnetocaloric properties well below room temperature. The RE$_{5}$Pd$_2$In$_4$ compounds with RE = Ho and Er are of special
interest as their $-\Delta S_{M}^{max}$, TEC, RC and RCP parameters reach the highest values when compared to those of remaining
RE$_{5}$Pd$_2$In$_4$ (RE = Tb, Dy and Tm).

\section*{Declaration of Competing Interest}

The authors declare no conflict of interest.

\section*{Acknowledgements}


The study was funded by ``Research support module'' as part of the ``Excellence Initiative -- Research University'' program
at the Jagiellonian University in Krak\'ow.

The research was carried out with the equipment purchased thanks to the financial support of the European Regional Development Fund
in the framework of the Polish Innovation Economy Operational Program (contract no.POIG.02.01.00-12-023/08).

	\bibliographystyle{elsarticle-num-names} 
	\bibliography{RE5Pd2In4_magn_effect_arXiv_20240226_AH.bib}
\end{document}